\documentclass[aps,prl,groupedaddress]{revtex4}

\begin{document}

\title{Reply to the Comment of M. V. Cheremisin (cond-mat/0408050)}

\author{S. A. Mikhailov}
\author{N. A. Savostianova}
\affiliation{Mid-Sweden University, ITM, Electronics Design Division, 851 70 Sundsvall, Sweden}

\date{\today}

\maketitle

In the preprint \cite{Cheremisin04b} M. V. Cheremisin noticed that some results (namely Fig. 4) of our recent work \cite{Mikhailov04b} have been taken from his paper \cite{Cheremisin04a}. As clear from the caption to Fig. 4 as well as from the text of our paper \cite{Mikhailov04b}, we did not consider the curves shown in that Figure as our original results, but emphasized that they have been obtained in the well known paper \cite{Chiu74} 30 years ago \cite{note2}. It is a surprise, that M. V. Cheremisin presents the 30-years-old results of Chiu and Quinn as his own results, obtained for the first time in \cite{Cheremisin04a}.


\begin{thebibliography}{1}
\bibitem{note2} Citation from the figure caption: ``The frequency versus magnetic field dependence of magnetoplasmon-polariton, as obtained in Ref.$^{13}$''; citation from the text: ``Figure 4 exhibits the calculated in$^{13}$ dependence of frequency on magnetic field'' (ref.$^{13}$ in \cite{Mikhailov04b} is the paper by Chiu and Quinn \cite{Chiu74}).
\end{thebibliography}
\end{document}